# Fractal Patterns in Music

John McDonough and Andrzej Herczyński

*Department of Physics, Boston College, MA 02467*

## Abstract

If our aesthetic preferences are affected by fractal geometry of nature, scaling regularities would be expected to appear in all art forms, including music. While a variety of statistical tools have been proposed to analyze time series in sound, no consensus has as yet emerged regarding the most meaningful measure of complexity in music, or how to discern fractal patterns in compositions in the first place. Here we offer a new approach based on self-similarity of the melodic lines recurring at various temporal scales. In contrast to the statistical analyses advanced in recent literature, the proposed method does not depend on averaging within time-windows and is distinctively *local*. The corresponding definition of the fractal dimension is based on the temporal scaling hierarchy and depends on the tonal contours of the musical motifs. The new concepts are tested on musical "renditions" of the Cantor Set and Koch Curve, and then applied to a number of carefully selected masterful compositions spanning five centuries of music making.

# I. Introduction

It has been long recognized that mathematics and music, despite their divergent modes of expression, are nevertheless "bound together," as Helmholtz put it.[1] The earliest exploration of this bond is usually credited to Pythagoras, but had likely been the work of his anonymous followers, the Pythagoreans, who assigned numerical values to notes plucked on λύρα – the lyre[2]. Among many mathematicians and physicists who contributed to music theory were Euclid, Ptolemy, Kepler, Huygens, Mersenne, Daniel Bernoulli, Fourier, and Euler. Jean-Phillip Rameau's 1722 *Treatise on Harmony* earned him the nickname "Isaac Newton of music." D'Alambert was so impressed by Rameau's mathematical principles of music that he wrote his own summary of the composer's theories. A beautiful historical outline is offered in R.C Archibald's 1924 essay[3]; a more recent account can be found in reference [4].

Perhaps the most fundamental link between the two fields lies in the construction of musical scales, a formidable challenge due to the logarithmic character of human (and not only human) hearing. The development of the equal tempered scales in use today has its own long and complicated history.[4-5] Instrument design and their acoustical properties depend, in turn, on the scales and involve physics principles and experimentation as much as art. Antonio Stradivari's artful innovations in perfecting string instruments, to take a celebrated example, set the standard followed meticulously to this day. The wave nature of sound, and sound propagation, are the domain of physics[6-9] as is the design of music halls, wherein art and physics meet.[10-13]

But beyond the physics of production, propagation, and reception of music, mathematics touches also the very form and structure of compositions, often providing the framework for pattern variations or suggesting the development of musical ideas. A number of common compositional stratagems are based on mathematical transformations of musical motifs, such as shifts, reflections ("vertical" in pitch or "horizontal" in time), elongations (stretches) and shortenings (compressions) – or in musical terms: transpositions, inversions, retrogrades, augmentations, and diminutions, respectively. Johann Sebastian Bach's the Art of the Fugue (*Kunst der Fuge*, BWV 1080) is an exquisite compendium of such techniques.

Bach's Contrapunctus VII, one of the 14 fugues[14] in the collection, deploys all these transformations and their combinations. The principal theme and its inversions appear in three different time scales: the original statement (*rectus*) or its inversion (*invertus*); the scaled-down version (diminution); and the scaled-up version (augmentation). These transformed motifs at different time-scales are layered in four voices (SATB) in intricate, and always changing kaleidoscopic ways, so that in some passages two of the lines of music are in the same tempo and the third is slower or faster, and at other times the theme unfolds on three different times scales simultaneously.

Such multi-scale, complex structures in music are reminiscent of visual fractals but rendered as time series of notes or chords—extended in time rather than in space. This paper aims to revisit the notion of scaling regularities in music and provide a more precise and more discriminating definition of the fractal pattern in sound based on this analogy, informed by a number of compositional techniques and the standard music-theoretical approaches.

## II. Review of Literature

It is remarkable that the literature on scaling regularities in music goes back at least to 1975, the same year that the first comprehensive treatise on fractals by Mandelbrot[15] was published. Perhaps the earliest contribution in this category were the papers by Voss and Clarke[16-17] who investigated the spectral density of audio power (effectively loudness) in a range of recordings, from news radio to jazz to classical music and found $1/f$ power law regularity present in all of them. This work stimulated many researchers to refine the method. In a variant, Hsu and Hsu[18] focused on note intervals and also found power law characteristics. Applied to music, spectral-densities may reveal scaling of the form $1/f^\beta$, indicating white noise ($\beta = 0$), pink noise ($\beta = 1$), red (or Brownian) noise ($\beta = 2$), and black noise ($\beta > 2$) characteristics. The index $\beta$ is a global measure, independent of details of the composition, a self-affine index rather than fractal dimension.[19] For a review and interpretation of $\beta$ see references [20-22]. Power laws have also been discerned in computer-generated "fractal" music.[23-25]

More recently, a number of studies applied the tools of statistical physics to selected musical pieces. Perhaps the most widely and successfully adopted method has been the detrended fluctuation analysis (DFA).[21, 26-28] The method is a refinement of the vector version of the fluctuation analysis designed to measure autocorrelation in multivariate time-series data. A thorough review of the literature on DFA, and the extension of this method to non-linear correlations, can be found in [21]. The authors identified a number of characteristic profiles in the fluctuation functions for a handful of (complete) compositions, from Palestrina to Shostakovich, and the corresponding Hurst exponents. A number of interesting classification possibilities and insights emerge from this approach. A similar statistical method, autocorrelation of normalized averages (ANA), was developed by Bigerelle and Iost[29] and used to classify music genres. Numerous other approaches based on statistical analysis, information theory, and Zipf's Law, have been developed.[30-34] A comprehensive review can be found in reference [30].

Statistical methods provide a valuable insight into the distribution of sequences of notes at various scales and may indeed reveal thematic autocorrelations across the composition, but they should be interpreted with caution. They are analogues of the usual box counting for patterns embedded in 2D, whereby "boxes" are replaced with time-windows. But this analogy entails a cognitive incongruity since an image (and any part of it) can be perceived in its entirety, whereas one can listen to a piece of music only one instant at a time, and it is impossible to hear it all at once. A composition of more than a few minutes may involve changes in tempo, sonority, loudness, and the very musical themes, and a listener's experience is not an amalgam of notes heard in *all* previous segments, of all possible durations, simultaneously. Music is localized in time, and held in memory with mental references to *selected* earlier episodes; it cannot be perceived, let alone comprehended, in an average sense.

Furthermore, all statistical analyses are insensitive to the order of the notes, and thus at a far remove from the meaning of music. A given composition rendered backwards in retrograde, would exhibit the same power laws for example, yet would likely amount to unintelligible noise (except for the rare pieces such as Crab Canon in the *Musical Offering,* BWV 1079, by Bach). Ignoring the order of notes would be akin to analyzing literature without concern for the order of the words. It is clear that a more satisfying approach is needed.

The autocorrelation, or Hurst exponent, obtained using either DFA or ANA method may not be related to the true fractal dimension unless the pattern exhibits self-similarity at progressively smaller scales. For example, in their analysis of Bach's sonata *sopr'il Soggetto Reale* (from *Musikalisches Opfer*, BWV 1079), González-Espinoza et. al.[21] indicate graphically that in the first movement, for which a single power law was discerned, a number of note patterns are repeated in each of the three voices (harpsichord, transverse flute, and violin). However, the repeated motifs appear always in the same tempo. What this suggests, then, is not self-similarity in the usual sense required of fractals, but the recurrence of its motivic elements at the same scale or autocorrelation. Consequently, González-Espinoza et. al.[21] do not identify their measured Hurst exponents with the fractal dimensions.

The standard box-counting algorithm, or its variants, can be used to analyze graphical representation of music[35-36] based on the pitch or loudness, creating a "musical contour." Chapter 7 of Charles Madden's book *Fractals in Music*[37] provides many examples with fractal dimensions in the range $1 < D < 2$. The principal advantage of this approach is its conceptual simplicity; the main shortcoming is that it renders (approximate) fractal dimensions without requiring any scaling structure in the music itself. This approach is critically assessed in a recent paper by Niklasson and Niklasson.[19]

A similar method for time-series, based on the variational principle, was developed by Dubuc et. al.[38] but as far as we know was not applied to music. Yet another box-counting approach uses scatter plots based on intervals between subsequent notes.[39-40] It is not clear why the resulting index can be interpreted as the fractal dimension. As the examples examined in chapter 6 of Charles Madden's book[37] illustrate, this index is not correlated to the complexity of the music. More fundamentally, these papers deploy a formula for fractal dimension, which is applicable only when the data conforms to a power law, which is manifestly not the case in these analyses.

A graphical approach of a different kind has been developed by James Walker and his collaborators, and used extensively in his book, Mathematics and Music[41] and a review paper.[42] It relies on discrete Gabor Transform with fixed (or variable) time windows to produce time-frequency plots called spectrograms (and also scalograms), which offer nuanced graphical "portraits" of musical pieces. Spectrograms can be used in a variety of effective ways but they are not meant to provide a single characteristic index of complexity or regularity.

In the present work, which evolved through close scrutiny of a large number of musical scores, from the Renaissance to the present time, the primary focus is on identifying self-similarity of particular compositions where such patterns are clearly present. Self-similarity in music is understood as recurrence of particular musical motifs, or their symmetric transformations, at various time scales. Here duration is analogous to length in visual fractals embedded in 2D, such as the Cantor Set and the Koch Curve (see, e.g., ref. [43-44]). This analogy is then extended further to offer a definition of the corresponding fractal dimension. The method offered here is simpler and more direct than statistical analyses developed heretofore; it is dictated by the patterns of notes and is distinctively *local*: it aims to discern specific passages or movements within larger musical pieces, which harbor more or less developed scaling regularities.

The proposed approach has three antecedents. The first is the general anticipation that only some, highly idiosyncratic music pieces can be construed as fractal patterns, whether they were conceived as formal explorations, as practiced by Bach, or incidental. Such interpretation has been adopted early by Claude Lévi-Strauss in his remarkable book *Look, Listen, Read*.[45] "The fractal character of musical composition" Lévi-Strauss wrote, "results from the fact that the relation between a small number of contiguous notes is repeated unchanged when those fragments are compared with more extensive passages from the same piece."[45]

Another crucial insight is due to Henderson-Sellers and Cooper[46] who proposed a definition of musical self-similarity based on note durations, analogous to the Cantor Set, specifying how the faster passages are to be related to the slower ones. This work, to our knowledge, provides the first precise definition of a musical fractal based on the development of a melodic line. The third illuminating insight is due to Brothers,[47] who suggests that in considering self-similarity of musical fragments it is necessary to distinguish a number of possible scaling regularities, such as duration scaling and structural scaling. In the approach presented here, these preliminary ides are extended to define what we term *nesting sequences*.

In what follows, a systematic analysis of self-similar patterns in sound is undertaken, aiming to synthesize insights from the approaches noted above. The paper is organized as follows. Section III proposes a new definition of fractal pattern in music. In Section IV the notion of temporal scaling is generalized. Section V introduces the concept of tonal (structural) complexity. Section VI offers an operative definition of the fractal dimension, and Sections VII-IX provide numerous examples taken from the music literature spanning over 500 years of music history. The paper closes with a summary in Section X and concluding remarks in Section XI.

## III. Musical Quasi-Fractals

There is currently no broadly accepted consensus of what constitutes a fractal pattern in music. Statistical analyses, such as those noted above, provide various measures of structural complexity for a sequence of notes, such as the self-affine index $\alpha$ or the autocorrelation (Hurst) exponent $H$, but these can be numerically computed for any time series without requiring any self-similarity, however loosely understood. Yet self-similarity across multiple scales, whether exact or approximate, is the hallmark of all fractals, their *sine qua non*. Further, the exponents $\alpha$ and $H$ can be understood to characterize asymptotically increasing periods, or long-time memory of the piece (whether perceived by a listener or not), whereas fractal dimension $D$ characterizes asymptotically decreasing periods, or instantaneous coincidence.

For a strictly self-similar time series, $D = 2 - H$, but in music that is highly improbable, except perhaps in very short passages. For whole multipart symphonic movements, whose character, motifs, instrumentation, scoring, and tempi, vary greatly, even approximate overall self–similarity is simply unachievable; indeed, it would hardly be desirable. Difficulties arise even with a single line of music since it is not a simple time series; notes have values (durations) and pitch (frequency) and both must be taken into account in defining self-similarity. Compositions made-up entirely from notes of the same pitches, or from notes of equal value, are rare, although such passages do occur sometimes in longer works.

The approach proposed here aims to avoid treating music as a simple time series, and to discern scaling regularity akin to algorithmically defined fractals, that is reoccurrence of the same pattern at finer and finer scales, with sensitivity to the melody itself. It is modeled on close analogy with infinitely fine mathematical fractals (such as the Cantor Set or the Koch Curve), and their quasi-fractal graphical representation wherein only a finite number of subscales can be manifest. In musical pieces, which are finite in length and (in the Western tradition) have limited range of note values, only a few scaling orders can be expected.

The key insight of this analogy is that whereas in a spatial quasi-fractal structure (such as an image), the same pattern recurs *simultaneously* at multiple spatial scales at a given location, in music the same melodic line must occur at multiple time scales at a given instant. A musical fractal pattern, therefore, requires that a motif be performed simultaneously at a few different tempos, creating an intricate interplay of the theme with its faster or slower versions – musical self-similarity.

Two specific compositional stratagems for realizing such complexities in music can be identified, both already suggested in theoretical literature. The primary musical genre, which depends on simultaneous rendition of the same theme at varied tempos, is the *prolation canon*, also called mensuration canon (H.J. Brothers[47] calls it "motivic scaling"). The technique was almost certainly invented by Johannes Ockeghem who also might have given it the name. Every movement of his remarkable *Missa Prolationum*, which survives in a manuscript completed in 1503, offers a different variant of the form. Ockeghem's most gifted student, Josquin des Prez, developed it further in his masses with an unmatched finesse and inventiveness.

The second fractal technique, described formally in its simplest version by Henderson-Sellers and Cooper[46], consists of sequential refinement of a given motif wherein each note is replaced by a faster rendition of the entire theme but transposed, so that that it starts with that very note. The melody is thus constructed by layering *nested sequences*, and encompasses simultaneity of multiple scales in a figurative sense, in a single line of music, which becomes faster and faster at each order, but retains the outline of the slower version in the leading notes of each scaled copy. This algorithmic technique has been deployed in creative ways by many composers and so merits inclusion here.

## IV. Temporal Scaling

The two forms of fractal patterns discussed above suggest that any viable definition of fractal dimension for a musical passage must discern a defining motif, or a theme, and should reflect both temporal scaling regularities and the tonal complexity of the theme. In this section we develop the concept of temporal scaling for both forms of fractal patterns in music, prolation canons and nested sequences, and the corresponding temporal fractal dimension, $D_o$.

Nesting sequences, wherein at each new order every note of the previous order is replaced by a copy of the motif in shorter notes (smaller time scale), can be understood as analogues of the Cantor Set construction. As such, this particular compositional method leads in the most natural way to the definition of $D_o$.

The form of nesting sequences was suggested first, in its simplest version, by Henderson-Sellers and Copper[46], with an example based on a four-note motif made of equal notes. In the second order, the theme is transformed into 16 quarter notes, then into 64 16th notes, etc., keeping the overall duration in each order the same. This is analogous to a version of Cantor-like construction wherein an interval of length 1 is replaced by four subintervals of length ¼, and so on, so that the number of copies at each new order increases by the factor $N = 4$ with the inverse scale $s = 1/\epsilon = (1/4)^{-1} = 4$, and thus $D_o = \ln(N)/\ln(s) = 1$.

It is therefore seen that the particular pattern offered in [46] is inescapably "Euclidean." To move beyond the integer dimension, duration of each new copy of the musical motif (scaling) must be incommensurate with the number of copies. As examples, consider musical analogues of the Cantor Set and the Koch Curve.

An exact musical replica of the Cantor Set would be unwieldy since durations of 1/3 of a given note value are not obtainable in a satisfying manner. The closest equivalent, but still cumbersome, would be to use the scaling factor of 3/8. A musically feasible variant is a pattern wherein each "interval" at a given order is replaced with two copies of ¼ length, so that

$$D_o = \frac{\ln(2)}{\ln(4)} = 0.5. \tag{4.1}$$

An example based on two-note and two-rests motif chosen to mimic the Cantor construction is given in Figure 1. The possibility of nested sequences with fractal dimension greater than one is a musical equivalent of the Koch Curve. Here again a variant is developed, wherein each note of the three-note motif is replaced by three notes of ½ duration. This gives

$$D_o = \frac{\ln(3)}{\ln(2)} \approx 1.58. \tag{4.2}$$

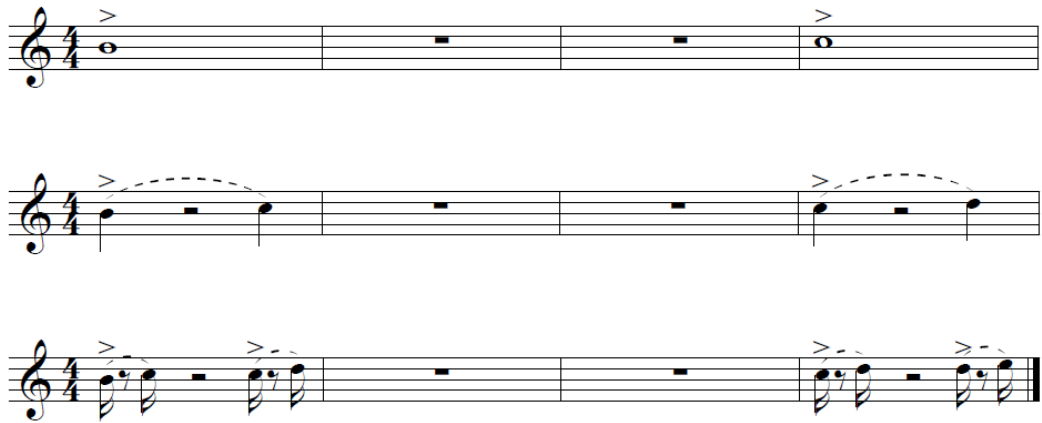

**Figure 1**. The Cantor-like nested sequences pattern based on a two note, two rests motif chosen for its visual similarity to a representation of the (special) Cantor Set. Each line of music corresponds to one order in scaling, with the overall duration (excluding rests) decreasing by ½ at each order. The carets indicate a note repeated (at shorter duration) from the previous order, and dashed slur brackets the entire motif at a given order.

Figure 2 provides an example with the three-note theme chosen to be visually similar to the graphical representation of the Koch Curve so that a triplet of notes, with the middle note raised relative to the first and the third, corresponds to a four-segment "hat" replacing straight segments in the lowest order of the Koch construction.

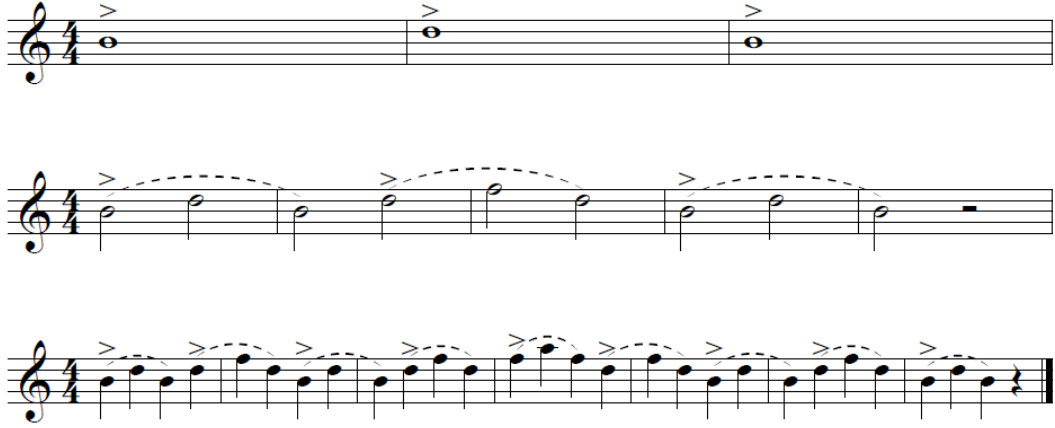

**Figure 2**. The Koch-like nested sequences pattern based on a three-note motif chosen for its visual similarity to a representation of the Koch Curve. Each line of music corresponds to one order in scaling, with the overall duration increasing by 3/2 at each new order. Carets indicate a note repeated (at shorter duration) from the previous order, and dashed slur brackets the entire motif at a given order.

It is thus clear that a non-integer scaling is possible with nested sequences, with the fractal dimension $0 < D_o < \infty$. Temporal scaling can be generalized further by allowing, at each order, notes of varied durations. For example, each whole note in the first order could be replaced by two eights and two quarter notes in the second and so on. In this case, the corresponding fractal dimension can be obtained using a more general approach[15] by setting

$$2 \left(\frac{1}{8}\right)^{D_o} + 2 \left(\frac{1}{4}\right)^{D_o} = 1, \tag{4.3}$$

or, denoting $\varphi = 2^{D_o}$,

$$\varphi^3 - 2\varphi - 2 = 0. \tag{4.4}$$

The cubic equation (4.4) has one real root, given, via the Cardano method, by

$$\varphi = (-4.5 + \delta)^{1/3} + (-4.5 - \delta)^{1/3}, \tag{4.5}$$

with $\delta = \sqrt{1 - 8/27}$, so that $\varphi \approx 1.77$ and

$$D_o = \frac{\ln(\varphi)}{\ln(2)} \approx 0.82. \tag{4.6}$$

As a complementary example with $D_o > 1$, consider a nesting sequence wherein in the second order each whole note is replaced by two quarter notes and two half notes, so that

$$2\left(\frac{1}{4}\right)^{D_o} + 2\left(\frac{1}{2}\right)^{D_o} = 1, \quad (4.7)$$

and again denoting $\varphi = 2^{D_o}$,

$$\varphi^2 - 2\varphi - 2 = 0. \quad (4.8)$$

or $\varphi = 1 + \sqrt{3}$, so that

$$D_o = \frac{\ln(\varphi)}{\ln(2)} \approx 1.44. \quad (4.9)$$

To extend this method of calculating $D_o$ to the *prolation canons*, consider first the simplest case, of two simultaneous voices proceeding at two different tempos, with the ratio of speeds of the first voice to the second given by a rational number $p/q$. If the two voices start and end together, such that each completes integer repeats of the defining motif, then each rendition of the theme in voice one is replaced by $p/q$ repeats in voice two, with $s = 1/\epsilon = q/p$, which implies $D_o = 1$. For $n$ simultaneous voices, all starting and ending together, it is thus sensible to define $D_o = n - 1$, each additional voice raising the temporal fractal dimension by 1.

In some prolation canons, however, separate voices do not start or finish together (or else do not fully complete repeats). Consider first canons with just two lines of music, and let $r_1$ and $r_2$ represent the number of repeats of the canon motif in the first and the second voice, respectively, where the two parameters are in general rational numbers but non necessarily integers. In that case, the number of scaled copies in the second line of music per one copy in the first line can be taken as $N = r_2/r_1$, so that

$$D_o = \frac{\ln(N)}{\ln(s)} = \frac{\ln(r_2/r_1)}{\ln(t_1/t_2)}, \quad (4.10)$$

where $t_1$ and $t_2$ are durations of the motif in the first and second voice, respectively, and the inverse time scale in the denominator is written as $s = q/p = t_1/t_2$. When the two voices start and finish at the same time, then $r_2/r_1 = t_1/t_2$, so that $D_o = 1$ as before.

Several properties of the definition (4.10) should be noted. First, indices "1" and "2" can be exchanged without changing the value of $D_o$, which means that the order in which the musical lines are written does not matter (as should be the case). Second, if $r_1 = r_2$ while $t_1 \neq t_2$ then $D_o = 0$, as expected. Finally, for $r_1 = r_2$ and $t_1 = t_2$, $D_o$ is indefinite; in this case the two voices are identical so that in effect there is just a single, doubled line of music and the notion of temporal scaling is simply not applicable.

For a prolation canon of $n$ voices, the relative scaling of each pair of voices must be included since the music depends on the simultaneous sounding of all such pairs. We therefore define

$$D_o = \frac{1}{n}\sum_{i\neq j}^{n} \frac{\ln(r_i/r_j)}{\ln(t_j/t_i)}, \qquad (4.11)$$

where the indices refer to the music lines' numbers. Here $N_{ij} = r_i/r_j$ is the number of copies in line $i$ relative to line $j$, and $\epsilon_{ij} = t_i/t_j$ is the scale of line $i$ relative to line $j$. Again, since in (4.11) any two indices can be exchanged without changing the value, the order of the music lines does not affect $D_o$. Equation (4.11) reduces to (4.10) for the prolation canon with two voices, or $n = 2$, as expected.

Definition (4.12) guarantees that when all pairs of voices are mutually commensurate, for example all voices start and end together and each repeats the motif integer number of times (a Euclidean canon, one might say), $D_o = n - 1$, in agreement with the earlier conjecture. Indeed, in this case all the terms under the sum in (4.11) are equal to 1 and so

$$D_o = \frac{1}{n}\left(2\binom{n}{2}\right) = n - 1. \qquad (4.12)$$

The two examples based on the Cantor Set and the Koch Curve in Fig. 1 and 2 can both be interpreted as prolation canons in three voices, taking each line to represent a separate, simultaneous voice. Thus, the two ways of realizing fractal patterns in music are seen as connected. It can then be immediately seen that for the Cantor Set Prolation Canon

$$D_o = \frac{1}{n}\left(2\frac{\ln(2)}{\ln(4)}\binom{n}{2}\right) = (n-1)D_c, \qquad (4.13)$$

where $D_c = \ln(2)/\ln(4) = 0.5$ is the fractal dimension of the Cantor-like structure. Similarly, a Koch Curve Prolation Canon can be constructed based on Figure 2, so that

$$D_o = \frac{1}{n}\left(2\frac{\ln(3)}{\ln(2)}\binom{n}{2}\right) = (n-1)D_K, \qquad (4.14)$$

where $D_K = \ln(3)/\ln(2) \approx 1.58$ is the fractal dimension of the Koch-like curve. Note that in neither case has the actual tonal content of the canonic motif been considered. We take up this aspect of musical fractal patterns in the next section.

## V. Tonal Complexity

The analogy between duration of motifs in compositions and spatial features of visual fractals serves as a useful guide for understanding temporal scaling in music, but it addresses only one of its aspects. Quintessential for the perception of the musical themes, and their role in compositional patterns, is their tonal character, the very melodic lines. Superposing a color scheme onto the Cantor Set could perhaps extend the analogy; in that case, the usual "black and white" Cantor Set would arguably correspond to musical motifs constructed from a series of notes of identical pitch, which would imply rather "colorless" compositions.

In order to account for pitch variations, a measure of tonal complexity is required, which would distinguish musical motifs of limited tonal variation from the passages incorporating larger or more frequent pitch intervals. We thus aim to define the parameter $x$ representing the tonal complexity of the following properties: (1) $x$ should be "local" and depend on the pitch intervals between neighboring notes; (2) the larger the interval the larger should be its contribution to $x$; (3) the larger the number of non-zero intervals, the larger the $x$; (4) a sequence of identical notes (no matter their duration), should yield $x = 0$ indicating the lowest possible tonal complexity (although not necessarily absence of any musical interest); and (5) a sequence with at least one non-zero interval should have $x > 0$.

In the twelve-note equal temperament scale, the relation between any two notes is defined by the ratio of their frequencies and is set to 2 for two notes an octave apart. For two notes, which are $j$ semitones apart, the ratio of their frequencies is thus given by,

$$\frac{f_{i+j}}{f_i} = \left(2^{1/12}\right)^j \approx \left(\frac{15.9}{15}\right)^j, \tag{5.1}$$

where the second, approximate expression is, in practice, used for tuning instruments. For the full octave, then

$$\frac{f_{i+12}}{f_i} = \left(\frac{15.9}{15}\right)^{12} = 2.0122 \approx 2, \tag{5.2}$$

as required. It follows that the intervals between notes should be measured in logarithms, as one would expect given that human hearing is logarithmic. This is, of course, the precise reason for the definition in (5.1). It is therefore convenient to define the tonal complexity parameter based on consecutive frequency intervals for an $N$-note motif as,

$$x = \frac{1}{\ln(2)} \sum_{i=1}^{N-1} \left|\ln\left(\frac{f_{i+1}}{f_i}\right)\right|. \tag{5.3}$$

The factor $\ln(2)$ is inserted to ensure that for any two notes an octave apart the corresponding tonal complexity is $x = 1$.

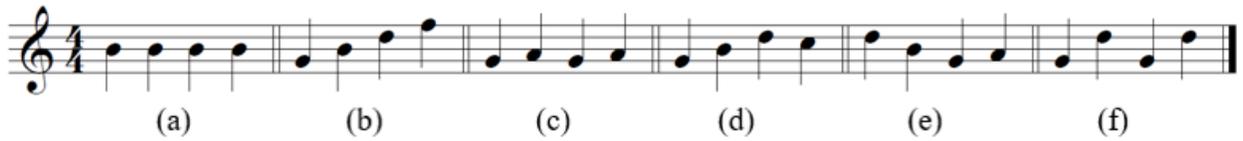

$$x_a = \frac{0}{12} + \frac{0}{12} + \frac{0}{12} = \frac{0}{12}, \quad x_b = \frac{4}{12} + \frac{3}{12} + \frac{3}{12} = \frac{10}{12}, \quad x_c = \frac{2}{12} + \frac{2}{12} + \frac{2}{12} = \frac{6}{12},$$
$$x_d = \frac{4}{12} + \frac{3}{12} + \frac{2}{12} = \frac{9}{12}, \quad x_e = \frac{3}{12} + \frac{4}{12} + \frac{2}{12} = \frac{9}{12}, \quad x_f = \frac{7}{12} + \frac{7}{12} + \frac{7}{12} = \frac{21}{12}$$

**Figure 3**. Six four-note example patterns and the corresponding $x$.

The parameter $x$ has all the required properties, some of which are illustrated in Figure 3. A sequence of notes of equal pitch gives $x = 0$, example (a); order of notes *does* matter, for example exchanging any pair of notes in example (b) would change $x$; inversion of a motif does not change $x$, examples (d) and (e); ascending and descending intervals contribute in the same way and only size of the intervals matters, examples (c) and (f). Finally, the retrograde version of any motif has the same complexity $x$ as the original theme.

For characterizing scaling properties, it is useful to introduce the normalized tonal complexity, effectively the average interval between consecutive notes, as

$$x_o = \frac{x}{N-1}. \tag{5.4}$$

Definition (5.4) ensures that simply repeating a given motif or its variant does not increase the overall complexity of musical compositions. Another useful property is that for a theme of $N$ notes ($N - 1$ intervals) all within one octave, the normalized complexity is no larger than 1. Indeed, using (5.3) and (5.4), the maximum possible $x$ in this case is

$$x = \frac{N-1}{\ln(2)} \ln\left(\frac{15.9}{15}\right)^{12} = N - 1, \tag{5.5}$$

so that $x_o = 1$. More generally, the maximum normalized complexity of a passage of arbitrary number of notes spanning $M$ octaves is $M$. It should be noted that these "Euclidean" dimensions correspond to the extreme cases when every interval is at the maximum, an unlikely musical feat. In Fig. 1, (a)-(f), $x_o = 0, 0.28, 0.17, 0.25, 0.25,$ and $0.58$, respectively.

## VI. Fractal Dimension

The parameter $D_o$ defined in Section III, provides a measure of temporal scaling regularity based on a musical motif, but regardless of its melody. The complexity index $x_o$ introduced in the previous section is a simple index of the tonal structure of a given fragment, independent of any scaling regularity. Together they can serve to capture salient features of quasi-fractal patterns in music. It is desirable to define the tonal dimension $D_t$ so that the overall dimension $D$ of a piece is given by the sum of the two, explicitly accounting for both the temporal and tonal aspects,

$$D = D_o + D_t . \tag{6.1}$$

Since $D$ must be zero in the absence of any scaling regularity, it is required that $D_t = 0$ when $D_o = 0$. Similarly, we expect $D_t = 0$ whenever $x_o = 0$, that is for a motif of notes equal in pitch as in Fig. 1(a). Finally, $D_t$ should be a monotonically increasing function of $x_o$. It follows that the tonal complexity should be understood as a factor modifying the scale so that $\epsilon' = \epsilon'(\epsilon, \chi_o)$ incorporates tonal characteristic. The simplest possible definition satisfying the conditions listed above is $\epsilon' = \epsilon \chi_o$, implying that the shorter the notes (higher order), the smaller the impact of the structural "weight" of the motif, consistent with an expected perception of a listener. Note that in order to guarantee that $\epsilon' < 1$, it is required that $x_o < 1$.

These considerations lead to

$$D_t = \frac{\ln(N)}{\ln(s')} = \frac{\ln(N)}{\ln(s(N-1)-\ln(\chi_0)}, \quad (6.2)$$

where $s' = 1/\epsilon'$. It is easy to verify that all the desirable properties listed above are satisfied by (6.2). In particular, for the set of six four-note examples shown in Fig. 1, and assuming the simplest case of temporal scaling when $D_o = 1$ or $\epsilon = 1/N$, as in the example considered by Henderson-Sellers and Cooper[46], equation (6.2) gives

$$\begin{aligned}D_{ta} &= 0, \quad D_{tb} = 0.52, \; D_{tc} = 0.44,\\ D_{td} &= 0.56, D_{te} = 0.56, D_{tf} = 0.72.\end{aligned} \quad (6.3)$$

To extend the definition of $D_t$ to prolation canons, the tonal complexity parameter $\chi_o$ must modify the temporal scales in the denominators of (4.11), $\epsilon_{ij} = t_i/t_j$, replacing them with $\epsilon'_{ij} = t_i\chi_o/t_j$. However, this procedure would violate the requirement that $\epsilon'_{ij} < 1$ for some of the elements in the sum, when $t_i > t_j$, so we proceed by setting

$$D_t = \frac{2}{n}\Sigma_{t_j>t_i} \frac{\ln(r_i/r_j)}{\ln(t_j/t_i)-\ln(\chi_0)}. \quad (6.4)$$

The definition (6.4) has two key proprieties. If $\chi_o = 0$ then $D_t = 0$, and if $\chi_o = 1$ then $D_t = D_o$. It is thus seen that $0 \leq D_t \leq D_o$ and therefore, based on (6.1),

$$D_o \leq D \leq 2D_o. \quad (6.5)$$

As an example, a three-voice prolation canon based on the melody given in Fig. 3(e) and Cantor-like temporal scaling of Fig. 1, would have $D_o = 1, \chi_o = 1/4$, and the tonal fractal dimension

$$D_t = \frac{2}{3}\left(2\frac{\ln(2)}{2\ln(2)+2\ln(2)} + 2\frac{\ln(2)}{4\ln(2)+2\ln(2)}\right) \approx 0.55, \quad (6.6)$$

so that the overall fractal dimension

$$D = D_o + D_t \approx 1.55. \quad (6.7)$$

The formalism developed above can be used to find a single, *effective* scaling parameter for any prolation canon, an additional index for these complex compositions. Following (6.2), the effective scale $\epsilon_o$ and the corresponding effective number of copies (repeats) $N_o$ of the canon theme for the *entire* composition must be related via

$$D_t = -\frac{\ln(N_o)}{\ln(\epsilon_o)+\ln(\chi_o)} \quad (6.8)$$

and from (4.10) also

$$D_o = -\frac{\ln(N_o)}{\ln(\epsilon_o)}. \tag{6.9}$$

With $D_o$ and $D_t$ given by (4.11) and (6.4), respectively, one can solve (6.8) and (6.9) for the effective parameters. This gives $\ln(N_o) = D_t D_o \ln(\chi_o)/(D_t - D_o)$ and

$$\epsilon_o = \chi_o^{D_t/(D_o - D_t)}. \tag{6.10}$$

Equation (6.10) requires that $\chi_o \neq 1$, consistent with the assumed condition $\chi_o < 1$. For the case of $\chi_o = 0$ and thus also $D_t = 0$, the effective scale $\epsilon_o$ is ill-defined. The effective scale $\epsilon_o$ is another index of a scaling pattern in music, one that can be expressed via the three parameters introduced earlier, $D_o$, $D_t$, and $\chi_o$, but which provides a succinct characterization of the overall scaling for a composition with multiple lines of music of arbitrary relative temporal scales. It is easy to verify that for a prolation canon of two voices, with commensurate tempos and the relative scale $\epsilon$, the effective scale is in fact the relative scale, $\epsilon = \epsilon_o$, and the effective number of copies is $N_o = 1/\epsilon_o$, as expected.

## VII. Examples: Prolation Canons

Prolation canons are the most direct, and arguably the most perfect realzations of a fractal patterns in music. They are the closest conceivable analogue of the spatial fractals, wherein the fine structure unfolds in time rather than in space, and self similarity is manifest across all the temporal orders present.

Like visual fractals, which may be defined by simple recursion relation and yet are "infinitely" complex, prolation canons are simple conceptually, but develishly hard to execute and therofore rare. Ockeghem might have develped the idea as an embelishement of a regualar canon, which was already pupular in the middle ages. In any case, the form is particualry suitable to multi-part choral pieces, and became popular in the Rennaisance. Prolation canons returned to favor in 20[th] century and a number of contempary composers have revived the form. Five examples, arranged chronolgically, are offerd below with detailed analyses.

**1. Johannes Ockeghem, *Missa Prolationum*: Kyrie eleison I (ca. 1450)**

This is a double prolation canon, wherein two separate pairs of voices, Soprano-Alto and Tenor-Bass, take up a different motif each. For each pair, the ratio of tempos is 3:2, with Soprano leading Alto, and Tenor leading Bass. A small complication is that the Bass line has a slightly different rhythm than the Tenor, and the speed ratio 3:2 is the average between these two voices. In the original *mensural notation* of Missa Prolationum, these relative tempos are ensured by the *mensuration* (analog of modern time signatures)[48].

The four parts begin simultaneously and the canon extends only through what is shown in Figure 4. The two faster voices proceed up to the asterisk and slower ones proceed through all three measures shown, so that all four follow the entire theme exactly once. After this passage, the music proceeds as a regular canon.

All movements of this mass harbor prolation canons, sometimes, as in Kyrie I, converting to a regular canon after a few bars. This device, wherein hidden complex structure leads to an accessible, transparent polyphony, is perhaps meant symbolically – as a transition from perplexity and doubt to enlightenment and clarity.

In each pair of voices, the theme consists of the six notes, performed once by each voice. The tonal complexity is $\chi_{SA} = 22/12 = 1.83$ for the Soprano-Alto pair, and $\chi_{TB} = 2$ for the Tenor-Bass pair so that $\chi_{oSA} = 0.37$ and $\chi_{oTB} = 0.40$ for the two pairs, respectively. Since the third bar in the first and third lines is not part of the canon, only the first two bars can be used to calculate fractal dimensions. The temporal dimension for both pairs of voices is thus $D_o = 1$. For the tonal dimensions, the sums in equation (6.4) will have just one element giving almost the same value for the SA and TB pairs, $D_{tSA} = 0.29$ and $D_{tTB} = 0.31$, and so $D_{SA} = 1.29$ and $D_{TB} = 1.31$. As a check of formula (6.10), notice that the two effective scales give the actual scales,

$$\epsilon_{oSA} = (0.37)^{0.41} \approx 0.66, \quad \epsilon_{oTB} = (0.4)^{0.43} \approx 0.66,$$

as expected.

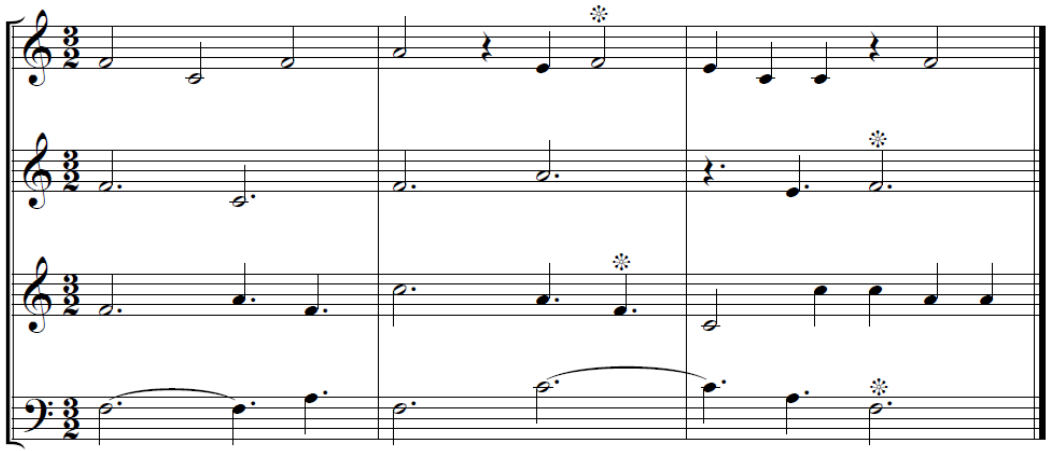

**Figure 4.** Ockeghem *Missa Prolationum Kyrie I*, the opening measures. The score is transcribed into modern notation with the asterisks inserted to indicate where each of the two motives ends in all four voices. The two faster voices, Soprano and Tenor (first and third line) lead and the Alto and Bass follow the two themes, respectively, at the slower tempo.

## 2. Josquin des Prez, Missa L'homme armé: Agnus Dei II (ca. 1495)

This is a supreme example of a pure prolation canon for three voices, Soprano, Alto and Tenor, with ratios of speeds 3:1:2, respectively, Figure 5. The three voices begin simultaneously, each following the same melodic line. The Soprano voice completes the entire melody, while the two slower voices go as far as the theme takes them at this time. The theme, as defined by the Soprano, extends over 25 measures and does not explicitly repeat earlier material. However, the motif given by the four opening notes (and its inversions) reoccurs often with slight variations. In the cadence this motif is interlocked with its own inversion, resolving to the final chord.

Based on the four-note motif, $\chi = 0.25$ whereas for the entire 68 note theme $\chi = 181/12 = 15.1$. Since the two values of normalized complexity differ by nearly the factor of 4, it is the one for the complete theme that must be used so that $\chi_o = 0.23$.

The numbers of repeats in the three lines (SAT) are $r_1 = 1$, $r_2 = 1/3$, and $r_3 = 1/2$, and the relative durations of the motif, as noted already, $t_1 = 1$, $t_2 = 3$, and $t_3 = 2$. It follows from (4.13) that $D_o = 2$, and $D_t = 0.64$ so that $D = 2.64$. The effective scaling for this three-voice canon is $\epsilon_o = 0.5$.

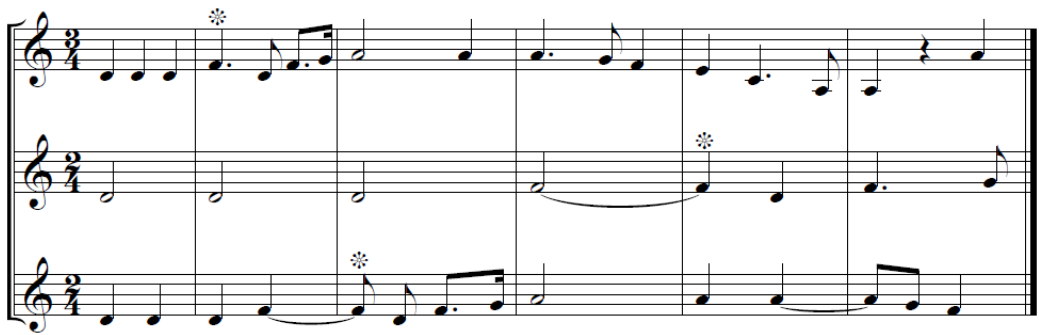

**Figure 5.** Josquin des Prez *Missa L'homme armé super voces musicales, Agnus Dei II*, the opening measures. Transcribed into modern notation using a different time signature for the soprano to clearly demonstrate the ratio of speeds, 3:1:2 for the first, second, and third lines, respectively. The asterisk indicates the for-note sub-motif, as rendered in the three voices, which serves as the basis for the entire theme defined by the Soprano.

### 3. J. S. Bach, Canon no. 14, BWV 1087 (after 1741)

The fourteen cannons in BWV 1087 were appended by Bach to his copy of the Goldberg Variations (BWV 988). The last of these, No. 14, is a prolation canon of a remarkably complex form, for four voices with the relative speeds 8:2:4:1 for Soprano, Alto, Tenor, and Bass, respectively. Figure 6 presents the five opening measures.

The four voices enter one after another; their starting delays are not proportional to their respective tempos. The main motif is given by the first eight notes (denoted by the solid slur bracket) in the soprano line, which is then repeated in a transposed version with the length of the last note slightly adjusted and followed by seven notes rephrasing the motif in inversion with some alterations (fragment with the dash slur bracket). The three lower voices each follow a different fragment of the soprano melody, so that the canon can unfold cyclically while maintaining the same relationship between each voice. The Alto follows the four first measures of the Soprano line, as shown in the figure. The Tenor repeats twice only the inverted fragment indicated by the dashed slur bracket, while the Bass that under the solid slur bracket – the original theme but inverted.

In the standard version, taken as the basis for calculation here, the Soprano line repeats the entire motif shown six times, but with an additional passage of 13-notes included in the middle (with the extra note appended at the end of the second bar).

Given that the complex structure is a based on an extension of the theme from the Goldberg Variations, these first eight notes can serve as the basis for calculating $\chi = 20/12 = 1.67$, and $\chi_o = 0.24$. Consistent with the outline given above, the number of repetitions for each of the four lines is set as $r_1 = 6, r_2 = 1, r_3 = 2$, and $r_4 = 1$. The durations of the motif are then taken as $t_1 = 0.5, t_2 = 2, t_3 = 1$, and $t_4 = 4$. These values give $D_o = 2.62$ and $D_t = 1.12$ so that $D = 3.74$. The effective scaling is $\epsilon_o = 0.32$.

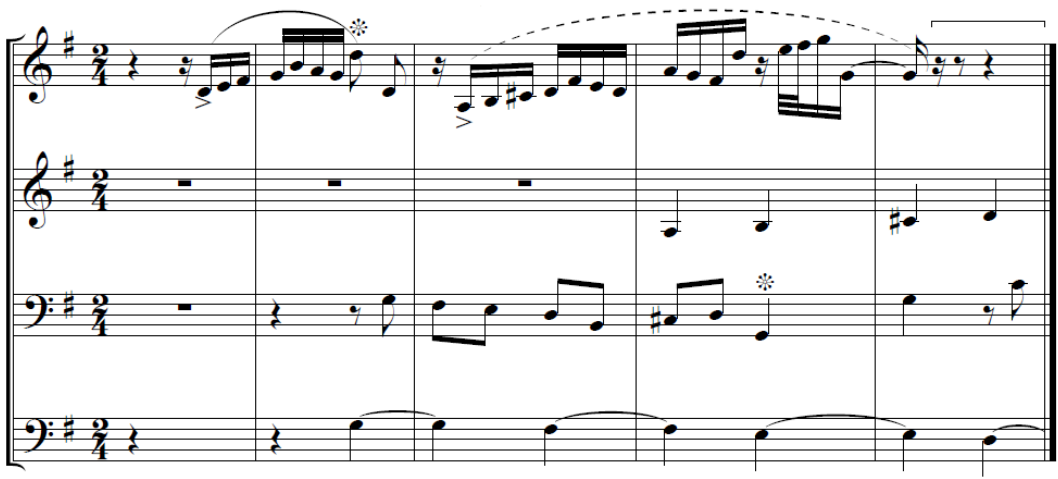

**Figure 6.** J.S. Bach *Canon no. 14 BWV 1087*, the opening measures transcribed to modern notation with the asterisks indicating the end of the main motif. The fragment shown in the Soprano line is followed by the Tenor in inversion; the one covered by the dashed slur bracket by the Alto; and the one covered by the solid slur bracket by the Bass in inversion. Rests covered by the bracket at the end of the Soprano line are replacing notes, excluded here for simplicity.

Canon No 7 (BWV 1087) is a simpler prolation canon, with the two upper voices following the Bass theme at a faster tempo. It should be noted that Bach wrote only a single line for each of these canons and did not indicate where the piece should end, so that they are "presented in the form of puzzles to be solved" as Martin Pearlman observed[49].

### 4. Johannes Brahms, *Schaffe in mir, Gott, ein rein Herz*, Op. 29 (1856-60)

The first movement of this motet for five voices (SATBB) is marked *Canon per augmentationem* indicating that this is a prolation canon with the ratio of speeds 2:1 for Soprano and Bass II, respectively. The Soprano repeats the 26-note theme twice, $r_1 = 2$, allowing the Bass II one full repeat (with omission of two of four consecutive F-naturals), $r_2 = 1$. The middle voices of the motet provide harmony and supporting counterpoint but do not participate in the canon. Here the complexity factor is $\chi = 50/12$ (it is not affected by the omission of two notes in the Base line) and so $\chi_o = 1/6, D_o = 1, D_t = 1.28$ so that $D = 2.28$. The effective scaling as given by (6.10) is $\epsilon_o = 0.5$, as expected since $t_2/t_1 = 2$.

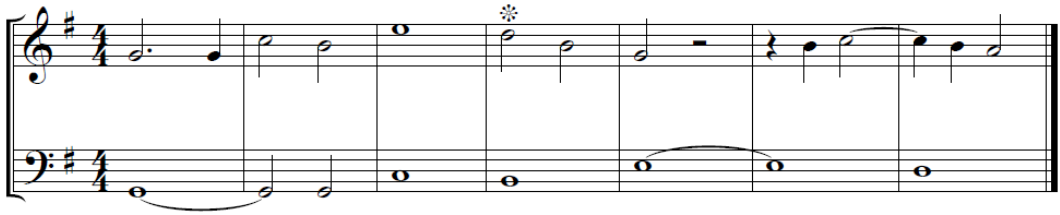

**Figure 7.** Brahms, motet Op. 29, *Schaffe in mir, Gott, ein Rein Herz,* opening bars (Andante moderato) showing the Soprano and Bass II voices (Alto, Tenor, and Bass I are omitted for clarity). The theme is stated in the Soprano in 12 bars, of which seven are shown. The Bass II line, at half the tempo, extends only through the part ending with the asterisk in the Soprano.

## 5. Dmitri Shostakovich, Symphony No. 15 in A Major Op. 141 (1971)

This prolation canon for three voices, in the first movement of the symphony, is the most intricate of the examples analyzed here. Shostakovich chooses an unusual ratio of tempos 8:6:5 for Violin I, Viola/Violin II, and Cello/Bass, respectively, leading to a highly complex counterpoint and self-correlation of the theme. In the transcription from the original score given in Figure 8, the proper relative tempos are ensured by three different time signatures[50].

The theme is based on an eight-note motif (A) indicated in Fig. 8 by the asterisk, but extends over 47 notes in the first line and has the structure AABAC, where B and C are two additional sub-motifs, made of 15 and 8 notes, respectively. The fastest voice follows the whole melody three times, $r_1 = 3$; the alto and the base enter at the beginning of the second and third repetition in the first line, respectively. Each voice continues to cycle at its own tempo and the canon ends with the slower voices cut off in the middle of the theme, so that $r_2 = 1.5$ and $r_3 = 0.5$ (strictly, $r_3 = 29/47$). The complex temporal structure gives unusual durations of the theme in each line, $t_1 = 6.0$, $t_2 = 7.92$, and $t_3 = 9.4$.

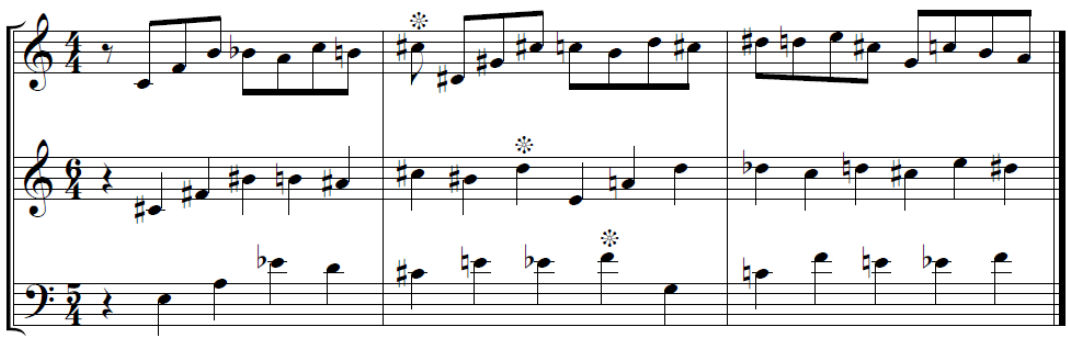

**Figure 8.** Shostakovich *Symphony No. 15 in A Major Op. 141,* Allegretto. The entrance of each voice in measures 255/267/273, respectively, is shown. The voices enter in order of the fastest to the slowest, first Soprano, then Alto, and finally Bass. The three voices, each transcribed into a single line from two, are displayed simultaneously to save space. Here, unlike in the original score, three different time signatures are used to clearly indicate different tempos.

The opening eight-note sub-theme has $\chi = 20/12 = 1.67$ and $\chi_o = 0.24$; for the entire 47-notes theme $\chi = 143/12 = 11.9$ and $\chi_o = 0.26$. Since the two values are almost the same, it is better to sue the one for the sub-theme as it is the basis for the entire melodic line. This gives $D_o = 8.55$, $D_t = 1.36$ and $\epsilon_o = 0.76$. The total fractal dimension, $D = 9.91$, is unusually high reflecting the intricate construction of this canon and the large number of repeats. Note that the remarkable character of the piece is due, in part, to the rests preceding the entrance of the second and the third voices, which are not shown in Fig. 8.

Among contemporary composers who adopted prolation canon to the modern idiom is Arvo Pärt. His remarkable *Cantus in Memoriam Benjamin Britten*, written in 1977, is a prolation canon for five voices, with the ratios of speeds 16:8:4:2:1 (Violin I, Violin II, Viola, Cello, Bass). The theme is based on the descending octave, developed sequentially with one note added at each repetition. The sequence, therefore, proceeds as A-A-G-A-G-F-A-G-F-E-A-… creating (in principle) ever-elongating fractal sequence.

## VIII. Examples: Nested Sequences

This technique, which has also been called "structural scaling" in a nomenclature proposed by Brothers[42], does not weave together separate voices proceeding at different tempos, as in a prolation canon. Instead, different temporal scales appear sequentially, in a single line of music, and at each order the entire passage is built from the copies of the original motif written in shorter notes. Each faster copy is transposed so that it begins with one of the notes of the previous order. In this manner, every new order retains the memory of all previous orders, in a progressively finer and self-similar structure. This algorithm is rather mechanical and highly prescriptive, so were it implemented rigorously, the resulting melody would appear deprived of spontaneity. It stand to reason, therefore, that the examples found in music literature depend on imaginative alteration, substitutions, and refinements, as illustrated by examples below.

**1. G. F. Handel, Suite No. 5 in E Major: *The Harmonious Blacksmith*, HWV 430 (1720)**

In this "transitional" example, the first and second order are interlocked with one another, so that it can be viewed as having simultaneous lines at different tempos, as in a prolation canon. Here, in measures 25-26, transcribed in Figure 9, a *double* nested sequence is developed to two orders. The motif, in the Bass line, is a three-note ascending stepwise melody rendered in two overlapping voices, one in quarter and another in eighth notes, denoted by the carets and the bars, respectively. In the second order (here simultaneous with the first), each note of the first order is replaced by a triplet-sixteenth notes outlining the original sequence. Similar interwoven structures at varied tempos abound in the composition, for example in a passage in variation 5 (measures 37-38) where each note of a four-note motif is reproduced in quartets of 32 notes. The three-note motif results in $\chi = 5/12$ or $\chi_o = 0.21$, and the two associated (simultaneous) time scales are $\epsilon_1 = 1/6$ and $\epsilon_2 = 1/3$. This leads to $D_o^1 = 0.61$, $D_o^2 = 1$, $D_t^1 = 0.33$, $D_t^2 = 0.41$, and finally the (total) fractal dimensions $D^1 = 0.94$ and $D^2 = 1.41$.

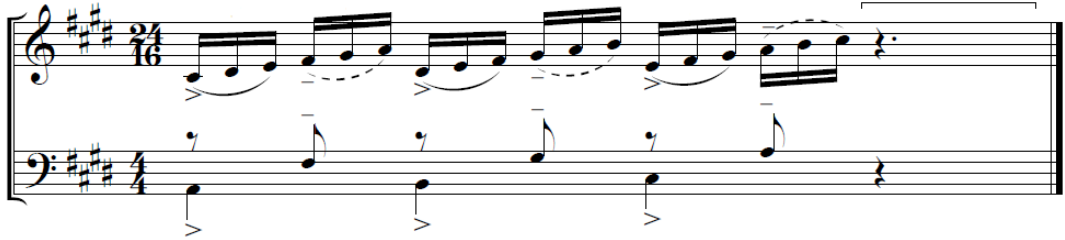

**Figure 9.** Handel *Suite No. 5 in E Major,* variation 3 (measures 25-26). The Bass line provides the first order, and the Soprano the second. Solid slur brackets and carets highlight the first motif, while dashed slur brackets and the bars indicate the second. The two motives are intertwined. Time signature for the right hand has been changed from the original score and the rests under the bracket at the end replace new musical material for clarity.

## 2. Joseph Haydn, Piano Sonata No. 53, Hob. XVI 34 (1783)

The first movement of this piano sonata has a remarkably regular form, with frequent repeats of a number of short motifs. Figure 10 displays a fragment (measures 98-101 and 103-106) based on a four-note theme, which is developed into a two-order nested sequence. This theme is an ascending four-note melody that outlines the low note of a sequence of arpeggios (three higher notes in each chord were omitted here for clarity). The first order consists of the Bass motif in doted half-notes for the left hand, while the second order, for the right hand (slightly altered), is inverted, appears delayed by one measure, and is rendered in the 16th. The temporal scaling factor is therefore $1/12$. Note that to add flair, Haydn included a fifth note to each of the "repeats," and also introduced a pair of leading notes followed by rests.

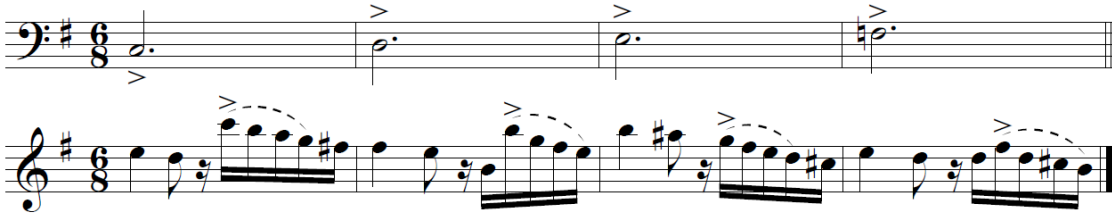

**Figure 10.** Haydn, Piano Sonata No. 53 (measures 98-101 followed by 103-106). The four-note motif is repeated numerous times (including in measure 102) but only the two fragments shown form the nested sequences. Carets indicate repeated (leading) notes and dashed slur bracket the entire inverted motif in the second order. The repeats in the second line are "dressed" with pairs of notes followed by the sixteenth rest, and one additional note is added in each.

Based on the four-note theme, $\chi = 5/12$ and $\chi_o = 0.14$. Since there are four copies in the second order at the scale $1/12$, it is seen that $D_o = 0.56$, $D_t = 0.29$ and so the total fractal dimension is $D = 0.85$.

### 3. Gustav Holst, *The Planets*: Uranus, Op. 32 (1917)

In the opening measures of Uranus by Holst, transcribed in Figure 11, nested sequences are developed to two almost complete orders. A four-note main theme is rendered first in dotted-whole notes, and then is repeated from the same starting note in dotted-quarter notes, with a longer fourth note. A second repetition casts it in dotted-eighth notes from the second note of the original theme; these two repetitions can be interpreted as a part of a second "asymmetric" order, wherein notes of unequal durations have been employed. Holst included only half of the second order (in the second line of music) so the nested sequence is incomplete.

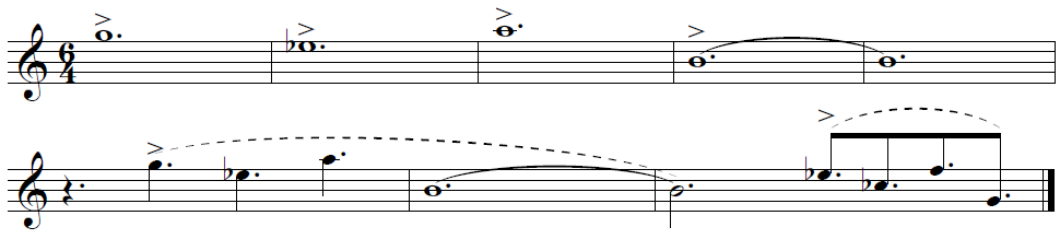

**Figure 11.** Holst *Uranus* from *The Planets*, Op. 32. The opening bars have been reduced to a single line and transposed up two octaves. Dotted notes replace bracketed tuplets in he original score to indicate the duration of each note directly. The C-flat note in measure 8 has replaced (as an enharmonic equivalent) the B note in the original to highlight the consistent melodic contour.

J.S. Bach, whose inventiveness and curiosity lead him to draw extensively on mathematical symmetries, also included nested sequences among his musical explorations. A notable example is his Sinfonia 15 from the *Three-Part Inventions* (BWV 801) for piano. Here, the theme, stated in the first measure in the right hand, consists of three sixteen-note triplets and is developed to two full orders in scales, with some inversions and harmonic alterations.

Another remarkable variant reminiscent of nested sequences, in Bach's Cello Suite No. 3, was noticed by Brothers[41]. The opening measures of Bourée I display "structural scaling" but only as far as the rhythmic pattern, AAB, is concerned. The actual music is altered at each subsequent iteration so that there is no singular motif, which reappears at various scales. This may be a very rare example (if not unique) of such temporal regularity.

## IX. Additional Examples

As might be expected, there are many instances in the music literature with elements of fractal writing, in a single voice or multi-part compositions, but wherein such developments are mathematically incomplete. In many cases, temporal scaling is present in fragmentary form, subsumed by unencumbered invention or abandoned mid-way to make an artistic statement. While it would not be informative to assign a fractal dimension to such passages, it may nevertheless be insightful to appreciate such mathematical symmetries as may be present.

One such example is Sonata in A Major, K. 268 (ca. 1750) by Domenico Scarlatti, with a passage that has characteristics of a nested sequence and simultaneously of a prolation canon, but neither developed fully. The fragment shown in Figure 12 is repeated four times in the piece. The starting note of each triplet follows the melody of the Bass notes denoted with carets. But here, the remaining notes of the Soprano line triplets do not follow the same theme, as would be expected for the nested sequences. The triplets serve as embellishments of the motif stated in the Bass line, so that the two versions interact as in a prolation canon.

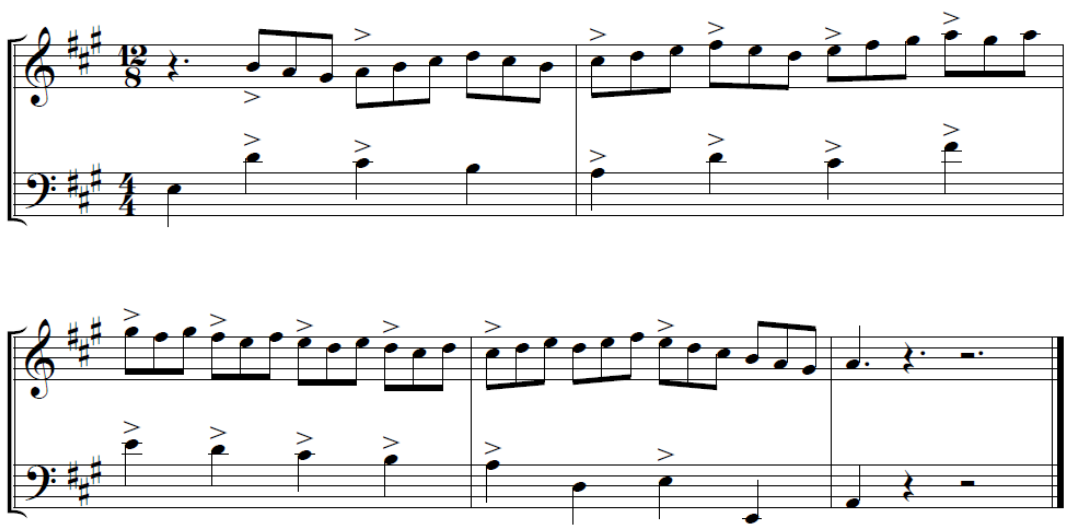

**Figure 12:** Scarlatti *Sonata in A Major K. 268,* bars 75-78, showing a quasi-fractal structure, wherein each quarter note marked with caret in the Bass corresponds to the first note of a triplet in the Soprano line. The faster (Soprano) line does not follow the nested sequence pattern. Time signature for the right hand has been changed from the original score for clarity.

A particularly intriguing example, from the mathematical point of view, is Johann Pachelbel's popular Canon in D. This is a regular canon for three voices, accompanied by an ostinato, with a very long theme made of ascending and descending stepwise melodies, which alter consistently every four bars. Since these structures reappear at various tempos, and are repeated with delays, they sound often simultaneously at two or three different speeds as in a prolation canon.

Per Nørgård is a contemporary composer who transcribes mathematically defined sequences of integers, "infinity series," into music. *Voyage into the Golden Screen* follows this model and is constructed so that the interval between consecutive even notes is fixed, and its inverse is the interval between consecutive odd notes. This intricate structure, however, is hard to perceive by the listener. The second movement is a prolation canon in 8 voices, where the ratio of tempos of the slowest and fastest line is a remarkable 1:32. Another contemporary composer, Conlon Nancarrow, took the mathematical structure of a prolation canon to a new level. He wrote a series of canons for player piano featuring odd tempo ratios between separate voices, practically impossible to play. In one of his canons, the ratio of speeds is set to $e/\pi \approx 0.865$; in another, 12 voices follow tempos based on ratios of frequencies in a just-intonation chromatic scale.

# X. Summary Discussion

The approach developed here aims to discern self-similar patterns in music recurring at various time scales. Two types of temporal scaling regularities can be identified in the music literature. In the prolation (or mensuration) canon, a number of separate voices proceed with the same motif simultaneously at different speeds, while in the nested sequences, finer time scales follow sequentially, but each iteration harbors all earlier versions. The two scaling modes are related, and in particular any nested sequence can be converted into a prolation canon.

The definitions of the temporal fractal dimension $D_o$ and the tonal fractal dimension $D_t$ accommodate both kinds of fractal patterns. The former reflects the rhythmic structure of the composition, whereas the latter reflects its melodic contour. The two attributes of music are then combined in a sum for the overall fractal dimension.

The interplay of rhythmic and melodic aspects of musical scaling regularities, reflected in the temporal and tonal fractal dimensions, makes the scaling analysis in music richer than its counterpart for the "black and white" spatial fractals. It is however limited otherwise since one can only expect a few orders of scales to be present in any composition. The range of note values (durations) is itself quite limited, and any composition encompassing more than three scaling orders would risk sounding overly mechanical, especially in the nesting sequences mode, or inscrutable in the prolation canon form.

Imperfections in musical scaling patterns derive also from the very nature of harmonies inherent in various musical tonalities. Algorithmic transformations, such as transpositions required of nesting sequences, may lead to harmonic inconsistencies or unexpected dissonances. Avoiding such harmonic irritants may require introducing small irregularities or reformulating the motif (Sinfonia 15 by Bach, BWV 801, can serve as an example). A composer may also wish to add an unexpected shift or an alternant such as an inversion of a small section or the entire theme, so that departures form the ideally conforming patterns are to be expected.

Among the prolation canons analyzed here, Shostakovich's from Symphony 15, with its unusual tempos, has by far the most complex temporal structure, with $D_o = 8.55$ and the total fractal dimension, $D = 9.91$. More typical structurally are prolation canons by Josquin, Bach and Brahms, with the fractal dimension in the range 2-3. Josquin clearly developed a much more technically daring approach than Ockeghem, whose double-canon while remarkably effective harmonically is characterized by a low dimension, $D \approx 1.3$. Even though these dimensions are only approximate indexes of scaling properties, they may nevertheless prove useful especially when compared with a listener's subjective experience of structural intricacies.

The compositions analyzed here were drawn from the oeuvre of the acknowledged masters in the history of music. They are not representative of that history, but show that fractal patterns have held unique fascination spanning over half a millennium. Our selective survey appears to suggest that mensuration canons were in vogue in the Renaissance, perhaps reaching their apogee with Josquin. They have enjoyed renewed popularity among contemporary composers. The nested sequences became fashionable among baroque masters and were arguably perfected by Bach, but can also be found in the classical period of the 18[th] century.

# XI. Concluding remarks

Can temporal scaling regularities be recognized while listening, and is it helpful to do so? Attention to fractal patterns, even if only fragmentary, may illuminate how the composition was crafted and enhance awareness of the development of musical narration. It is perfectly possible, of course, to enjoy the *Art of the Fugue* without noticing the intricate transformations of its singular subject. And yet following the imitative counterpoint is rewarding and enriches the listening experience – all the more so when one can contrast and differentiate rhythmic and tonal patterns within these variations.

Another question one may raise is whether mathematical regularities present in music are always intentional. That is surely the case in all prolation canons, whose intricate sound tapestry must be conceived *a priori*. It is possible that some approximate nesting sequences (perhaps that of Handel in Fig. 9) arose more or less spontaneously; if so, by following what they knew would sound natural and effortless, their composers intuitively arrived at multiscale patterns.

Temporal scaling in broader sense can be found not only in classical repertoire but also in folk and traditional music. In Indonesia, gamelan ensembles feature a unique type of polyphony called colostomy, wherein simultaneous lines of music, performed on metallophones and other percussion instruments, are nested with commensurate ratios of tempos. While this technique involves a rhythmic interplay of melodies played simultaneously on different instruments, they do not usually share any motivic elements.

Structures, which exhibit approximate self-similarity extending over a few orders in scales are ubiquitous in Nature, in the forms of plants, trees, and mountain contours, accounting perhaps for our esthetic preference for them in visual art. Quasi-fractal patterns in music may be appealing for similar reasons, even when they appear fleetingly. They convey a sense of graceful development, as if each new note found its natural, anticipated place.

We end with a cautionary note. Levi-Strauss, who was perhaps the first to describe what fractal scaling in music could mean, was also dismissive of it as a technique "from which one should not expect anything more than a tolerable acoustic ambiance."[45] Despite his enthusiasm for fractals in art and their "refinement and complexity,"[45] Levi-Strauss warned that "fractal algorithms do not have the ability to engender, whether in painting or music, more than those minor genres I have called decorative […]. A large gap separates these often fascinating objects from an authentic painting or a piece of music."[45]

Yet scaling regularities in music, as understood and illustrated here, may defy this early assessment. When deployed sparingly, accompanied by other musical devices, or infused with artistic license, fractal patterns may add richness, texture, and a touch of inevitability to the piece. Occasional subversion of mathematical symmetries can guard the composition from sounding overly prescribed or mechanical. The masterpieces analyzed here, as so many other examples in music literature, attest to the sovereignty of taste over technique: even the most refined musical formula must be "well-tempered" for the ear.

## Acknowledgments

During the course of work on this paper we benefitted from discussions with a number of musicians who were extraordinarily generous with their time and advice. We would especially like to thank Peter Phillips for sharing his insights regarding the use of mensuration canons in the Renaissance, and to the choir he directs, *The Tallis Scholars*, for their masterful recordings of all masses by Josquin, on which we relied. We are grateful to Steven Lipsitt, music director of *Bach, Beethoven, & Brahms* Orchestra, for alerting us to Brahms' motet Op. 29, and to Alyssa Wang, founder and artistic director of *Boston Festival Orchestra*, for her careful reading of the manuscript and sharing many discerning ideas and observations. Thanks are also due François Bergeron, for his splendid lecture on symmetry in Bach's music as well as his comments in our subsequent correspondence. We are grateful to Ann Lucas for pointing out canonic structures in non-Western music, particularly the colotomic rhythms in Indonesian music, and to Grae Worster, for his helpful questions and recommendations. Finally, AH would like to thank Burt Howell and Boston College's Intersections Program, for the invitation to a writing retreat in June 2022, where substantial portions of this paper were written.

# References

1. H. von Helmholtz, Vorträge und Reden, vol. 1, 82, Braunschwieg, 1884.
2. S. Weinberg, To explain the world: the discovery of modern science, Ch. 2, Harper Colllins Publishers, 2015.
3. R. C. Archibald, Mathematicians and music, Am. Math. Monthly, 31 No. 1, 1-25 (1924).
4. J. Fauvel, R. Flood, and R. Wilson, Eds., Music and mathematics: from Pythagoras to fractals, Oxford University Press, 2003.
5. J. S. Walker and G. W. Don, Mathematics and Music, 2nd ed., CRC Press, 2020.
6. P. M. Morse, Vibration and sound, Acoust. Soc. of Am., 1936.
7. J. Bachus, The Acoustical Foundations of Music, Norton, 1969.
8. T. D. Rossing, F. R. Moore, and P. A. Wheeler, The science of sound, Addison Wesley, 3rd ed., 2002.
9. J. S. Rigden, Physics and the sound of music, 2nd ed., 1985.
10. M. R. Schroeder, Computer models of concert hall acoustics, Am. J. Phys 41, 461-471 (1973).
11. M. R. Schroeder, Concert Halls, in The Psychology of Music, Elsevier 1999, 25-46.
12. D. Howard and L. Moretti, *Sound and space in Renaissance Venice: architecture, music, acoustics*, Yale University Press, 2010.
13. L. Beranek, How they sound: Concert and opera halls, Acoust. Soc. of Am., 1996.
14. F. Bergeron, Bach and the mathematics of the fugue (private communication).
15. B. B. Mandelbrot, The fractal geometry of nature, W. H. Freeman and Co. (1982).
16. R. F. Voss and J. Clarke, 1/*f* Noise in Music and Speech, *Nature*, 258, 317-318 (1975).
17. R. F. Voss and J. Clarke, 1/f noise in music: music from 1/*f* noise, J. Acoust. Soc. Am., 63 No 1, 258-163 (1978).
18. K. J. Hsu and A. J. Hsu. Fractal Geometry of Music, *Proc. Natl. Acad. Sci. U.S.A.* 87, No. 3, 938-941 (1990).
19. M. H. Niklasson and G. A. Niklasson, The fractal dimension of music: melodic contours and time series of pitch, arXiv: 2004.02612 (2020).
20. M. Gardner, Fractal music, hypercards and more: Mathematical Recreations from Scientific American Magazine, Chapter 1, Freeman and Company, 1991.
21. A. González-Espinoza, H. Larralde, G. Martinez-Mekler, and Markus Müller, Multiple scaling behavior and nonlinear traits in music scores, R. Soc. Open Sci. 4, 171282 (2017).
22. D.J. Levitin, P. Chorda, and V. Menon, Musical rhythm spectra from Bach to Joplin obey a 1 power law, PNAS, 109 No. 10, 3716-3720 (2012).
23. P. Campbell, Nature, 324, 523-528 (1986).
24. M. R. Schroeder, Is there such a thing as fractal music, Nature 325, 765-766 (1987)
25. P. Campbell, Nature 325, 766 (1987).
26. S. Sanyal, A. Banerjee, A. Patranabis, K. Banerjee, R. Sengupta, and D. Ghosh, A study of improvisation in a musical performance using multifractal detrended cross correlation analysis, Physica A, 462, 67-83 (2016).
27. L. Telesca and M. Lovallo, Revealing competitive behaviours in music by means of the multifractal detrended fluctuation analysis: application to Bach Sinfornias, Proc. R. Soc. A, 467, 3022-3032 (2011).
28. T. C. Roeske, D. Kelty-Stephen, and S. Wallot, Multifractal analysis reveals music-like dynamics structure in songbird rhythms. Sci. Reports, 8, 4570 (2018).

29. M. Bigerelle and A Iost, Fractal; dimension of music and classification of music, Chaos, Solitons, and zfractals, 11, 2179-2192 (2000).
30. J. Useche and R. Hurtado, Melodies as maximally disordered systems under macroscopic constraints with musical meaning, Entropy, 21, 532 (2019).
31. A. Aydemir and G. Gündüz. Fractal dimensions and entropies of Meragi songs, Ch. 6 in Nonlinear Phenomena in Complex Systems: from Nano to Macro Scale, D. Matrasulov and H. E. Stanley, Springer 2014.
32. S. Nicholson and E. Kim, Structures in Sound: Analysis of classical music using the information length, Entropy 18, 258 (2016).
33. D. Adams and P. Grigolini, Music, New Aesthetic, and Complexity, in Complex 2009, Part II, LNICST 5, J. Zhou (Ed.), 2212-2221 (2009).
34. A. Pace, K. Mahmoodi, B.J. West, Complexity measures in music, Chaos, Solitons and Fractals, 108, 82-86 (2018).
35. A. Zlatintsi and P.Maragos, Multiscale fractal analysis of musical instrument signals with application to recognition. IEEE Trans. Audio Speech and Language Proc., 21 No 4, 737-748 (2013).
36. Zhi-Yuan Su and Tzuyin Wu, Multifractal analyses of music sequences. Physics D, 221, 188-194 (2006).
37. C. Madden, Fractals in music, 2$^{nd}$. ed., High Art Press 2007.
38. B. Dubuc, J. F. Quiniou, C. Roques-Carmes, C. Tricot, and S. W. Zucker, Evaluating the fractal dimension of profiles, Phys. Rev. A, 39 No. 3, 1500-1512 (1989).
39. G. Gündüz and U. Gündüz, The mathematical analysis of some songs, Physica A, 357, 565-592 (2005).
40. A. Georgaki and T. Christos, Fractal based curves in musical creativity: a critical annotation, in: Chaos Theory: Modeling, Simulation and Applications, 167-174, World Scientific 2011.
41. G. W. Don, K. K. Muir, G. B. Volk, J. S. Walker, Music: broken symmetry, Geometry, and complexity, Notices of the AMS, 57 No 1, 30-49 (2010).
42. J. F. Alm and J. S. Walker, Time-frequency analysis of musical instruments, SIAM Review, 44 No. 3, 457-476.
43. S. H. Strogatz, Nonlinear Dynamics and Chaos, Perseus Books, 1994.
44. M. F. Barnsley, Fractals Everywhere, Morgan Kaufmann, 2000.
45. C. *Lévi*-Strauss, Look, listen, read, Basic Books, 1997 (translation of "Regarder, Écouter, Lire," Librairie Plon, 1997).
46. B. Henderson-Sellers and D. Cooper, Has Classical Music a Fractal Nature?: A Reanalysis. *Computers and the Humanities.* 27, No. 4, 277-284 (1994).
47. H. J. Brothers, Structural Scaling in Bach's Cello Suite No. 3. *Fractals*, 15 No. 1, 89-95 (2007).
48. Scott Metcalfe, program notes for the Blue Heron ensemble performance of Ockeghem's *Missa Prolationum*, March 9, 2019, Frist Congregational Church, Cambridge, MA.
49. M. Pearlman, program notes for the Boston Baroque Orchestra performance of "Goldberg" Canons, BWV 1087, Isabella Gardner Museum, January 16, 1990.
50. In the original Shostakovich score, with the single time signature, the two slower voices begin at the second and third repetition of the theme of the fastest line. The alto voice enters 12 measures after the soprano, and the third 6 measures later. This is possible since the signature changes from 2/4 to 4/4 when the second voice enters.